\documentclass[12pt]{article}
\usepackage{epsfig}
\begin{document}
\title{Antikaon-induced $K_1(1270)^+$ meson production on nuclei near threshold}
\author{E. Ya. Paryev$^{1,2}$\\
{\it $^1$Institute for Nuclear Research, Russian Academy of Sciences,}\\
{\it Moscow 117312, Russia}\\
{\it $^2$Institute for Theoretical and Experimental Physics,}\\
{\it Moscow 117218, Russia}}

\renewcommand{\today}{}
\maketitle

\begin{abstract}
We study the inclusive strange axial-vector meson $K_1(1270)^+$ production in ${K^-}A$ reactions at near-threshold
laboratory incident antikaon momenta within a nuclear spectral function approach,
which describes incoherent direct $K_1(1270)^+$ meson production in $K^-$ meson--proton
${K^-}p \to {K_1(1270)^+}\Xi^-$ production processes and accounts for three different options for its
in-medium mass shift (or for its effective scalar potential) at central density $\rho_0$.
We calculate the absolute differential and total cross
sections for the production of $K_1(1270)^+$ mesons on $^{12}$C and $^{184}$W target nuclei at
laboratory angles of 0$^{\circ}$--45$^{\circ}$ by $K^-$ mesons with momenta of 2.5, 2.8 and 3.5 GeV/c,
which are close to the threshold momentum ($\approx$ 2.95 GeV/c) for $K_1(1270)^+$ meson production off the free
target proton at rest. The intrinsic properties of carbon and tungsten target nuclei have been described in terms
of their spectral functions, which take into account the momenta of target protons and the energies of
their separation from the considered nuclei.
We show that the differential and total (absolute and relative) $K_1(1270)^+$ antikaon-induced
production cross sections at initial momenta not far from threshold -- at momenta $\sim$ 2.8--3.5 GeV/c, at which
there is yet no strong drop in their strength, reveal
a distinct sensitivity to changes in the in-medium shift of the $K_1(1270)^+$ mass, studied in the paper,
both in the $K_1(1270)^+$ meson low-momentum (0.1--1.0 GeV/c) and in its full-momentum ranges.
This would permit evaluating this shift. Experimental data necessary for this aim can be obtained
in a dedicated experiment at the J-PARC Hadron Experimental Facility.
\end{abstract}

\newpage

\section*{1 Introduction}

\hspace{0.5cm} The study of the in-medium properties (masses and widths) of hadrons at finite density
has received considerable interest in the last two decades due to the hope to extract valuable information
on the partial restoration of chiral symmetry in a nuclear medium (see, for example, [1--6]).
Another prominent case of medium modification of hadrons is that of the strange vector $K^*(892)$
and axial-vector $K_1(1270)$ mesons with the same charge states (or with the same quark structure
$q{\bar s}$ or ${\bar q}s$ with $q=u,d$), which are chiral partners and
whose in-medium mass difference, as is expected [7--10], is sensitive to the chiral order parameter and,
hence, will give the possibility to identify unambiguously the effect of chiral symmetry
breaking in nuclear medium. Their isospins, spins-parities quantum numbers are
$I(J^P)=\frac{1}{2}(1^-)$ for the $K^*(892)$ and $I(J^P)=\frac{1}{2}(1^+)$ for the $K_1(1270)$.
They have a relatively small vacuum decay widths of 50 and 90 MeV, respectively.

 At present, in literature there is a number of publications devoted to the study of the in-medium
properties  of hadronic resonances $K^*(892)$ and $K_1(1270)$. The properties of ${\bar K}^*(892)$ and
$K^*(892)$ mesons in cold nuclear matter have been studied in Refs. [11--13] and [13, 14], respectively,
basing on chirally motivated model of the meson selfenergies.
Thus, for instance, a moderately attractive and repulsive real
low-energy nuclear potentials (or their in-medium mass shifts) of about -50 and +40 MeV at density $\rho_0$
have been predicted for the ${\bar K}^*(892)$ and $K^*(892)$ mesons, respectively.
These are similar to those for light strange mesons ${\bar K}$ and $K$.
On the other hand, a negative mass
shift of about -20 MeV has been predicted for the $K^*(892)^+$
meson, having the same quark content $u{\bar s}$ as the $K^+$ one, at rest at saturation density $\rho_0$
within the quark-meson coupling model [15].
Other recent calculations [8] based on QCD sum rule predict
that the upper limits of the mass shifts of $K_1^-$ and $K_1^+$ mesons in nuclear matter are
-249 and -35 MeV, respectively.

The influence of the $K^*(892)^+$ meson in-medium mass shift on its yield from ${\pi^-}A$ reactions
at near-threshold laboratory initial pion momenta of 1.4--2.0 GeV/c has been recently analyzed within
the collision model [16] using an eikonal approximation. It has been shown that both differential and
total $K^*(892)^+$ production cross sections possess a high sensitivity to changes in this shift
in the low-momentum region of 0.1--0.6 GeV/c.
In addition to a feasibility study performed in Ref. [16] in order to investigate the impact
of the mass shift of $K^*(892)^+$ mesons in nuclear matter on their yield in near-threshold ${\pi^-}A$
reactions, we give here the predictions
for the absolute differential and total cross sections for near-threshold production of $K_1(1270)^+$ mesons in
${K^-}^{12}{\rm C} \to {K_1(1270)^+}X$ and ${K^-}^{184}{\rm W} \to {K_1(1270)^+}X$ reactions at laboratory angles of
0$^{\circ}$--45$^{\circ}$ by incident antikaons with momenta of 2.5--3.5 GeV/c as well as for their relative yields
from these reactions within three different scenarios for the $K_1(1270)^+$ in-medium mass shift.
These nuclear targets were also adopted in calculations [16]
of $K^*(892)^+$ meson production in ${\pi^-}A$ reactions
and, therefore, can be used in studying the ${K^-}A \to {K_1(1270)^+}X$ interactions here.
The present calculations are based on a first-collision model, developed in Ref. [16] for the
description of the inclusive $K^*(892)^+$ meson production
and extended to account for different scenarios for the $K_1(1270)^+$ in-medium mass shift and width change.
The medium modification of the $K_1(1270)^{\pm}$ (and $K^*(892)^{\pm}$) mesons could be probed
at the J-PARC Hadron Experimental Facility through the $K^-$ reaction on various nuclear targets [8]
\footnote{$^)$It is appropriate to note that the measuring of $K_1(1270)^-$ and $K^*(892)^-$ mesons would
require less energy to produce compared to that for $K_1(1270)^+$ and $K^*(892)^+$ using the $K^-$ beam
due to more lower their lowest production thresholds in elementary reactions
$K^-p \to K_1(1270)^-p$ and $K^-p \to K^*(892)^-p$ ($\approx$ 1.94 and 1.08 GeV/c, respectively)
compared to those for $K_1(1270)^+$ and $K^*(892)^+$ production in the relevant elementary channels
$K^-p \to K_1(1270)^+\Xi^-$ and $K^-p \to K^*(892)^+\Xi^-$ ($\approx$ 2.95 and 1.95 GeV/c, correspondingly).
Therefore, such experimental possibility may be provided for them already now by the available at J-PARC
high-intensity separated secondary antikaon beam with the momentum up to 2 GeV/c. In this connection, it
is interesting to extend in the future the study of Ref. [16] and of present one also to the consideration
of $K_1(1270)^-$ and $K^*(892)^-$ production on nuclei in near-threshold $K^-A$ interactions to get a
deeper insight into the possibility of observing here of their modification in nuclear medium in these
interactions.}$^)$
.
Comparison the results of such measurement for $K_1(1270)^+$
and those from $K^*(892)^+$ measurement in ${\pi^-}A$ reactions with our
calculations -- previous [16] and present -- will shed light on the
partial restoration of chiral symmetry in nuclear matter [8].

\section*{2 Framework: direct process of $K_1(1270)^+$ meson production in nuclei}

\hspace{0.5cm} Direct production of $K_1(1270)^+$ mesons in $K^-A$ (A=$^{12}$C and $^{184}$W)
interactions at near-threshold incident antikaon beam momenta below 3.5 GeV/c may occur
in the following $K^-p$ elementary process, which has
the lowest free production threshold momentum ($\approx$ 2.95 GeV/c):
\begin{equation}
K^-+p \to K_1(1270)^++\Xi^-.
\end{equation}
We can ignore in the momentum domain of interest the contribution to the $K_1(1270)^+$
yield from the processes $K^-p \to {K_1(1270)^+}\Xi^-{\pi^0}$, $K^-p \to {K_1(1270)^+}\Xi^0{\pi^-}$,
$K^-n \to {K_1(1270)^+}\Xi^-{\pi^-}$ and $K^-p \to {K_1(1270)^+}\Xi(1530)^-$
due to larger their production thresholds ($\approx$ 3.33 and 3.57 GeV/c, respectively) in $K^-N$ collisions.

Following [16], we simplify the ensuing calculations via accounting for the in-medium modification
of the final $K_1(1270)^+$ meson, involved in the production process (1), in terms of its
average in-medium mass
$<m_{K_1}^*>$ instead of its local effective mass $m_{K_1}^*(|{\bf r}|)$ in the in-medium cross section
of this process, with $<m_{K_1}^*>$ defined as:
\begin{equation}
<m^*_{K_1}>=m_{K_1}+V_0\frac{<{\rho_N}>}{{\rho_0}},
\end{equation}
where $m_{K_1}$ is the $K_1(1270)^+$ free space mass, $V_0$ is the $K_1(1270)^+$ effective scalar
nuclear potential (or its in-medium mass shift) at normal nuclear matter
density ${\rho_0}$, and $<{\rho_N}>$ is the average nucleon density.
In the present work, for nuclei $^{12}$C and $^{184}$W of interest,  the ratio $<{\rho_N}>/{\rho_0}$,
was chosen as 0.55 and 0.76, respectively. No experimental data exist presently on the $K_1(1270)^+N$
interaction. Recent calculations performed in Ref. [8] within the QCD
sum rule approach show, as was noted above, that the upper limit of the mass shift of $K_1(1270)^+$ mesons
in nuclear matter is -35 MeV at zero momentum relative to the surrounding nuclear matter.
Therefore, we will adopt for the quantity $V_0$ in this work
three following options: i) $V_0=-35$ MeV, ii) $V_0=-20$ MeV and iii) $V_0=0$ MeV.
Since there is no information about momentum dependence of this mass shift, we will use the above options for it
at finite momenta, accessible in calculation of the $K_1(1270)^+$ production in $K^-A$ reactions at beam momenta of
interest, with allowance for the Fermi motion of intranuclear protons (at momenta $\sim$ 0.1--2.7 GeV/c, see below).
The experimental knowledge on the $\Xi^-N$ interaction is poor currently. The study of $\Xi^-$ hypernuclei provides
a valuable information on this interaction at low energies [6, 17]. At present, there are only a few sets of
experimental data on the $\Xi^-$ hyperon properties in the nuclear medium.
Following phenomenological information deduced in Refs. [18, 19] from old emulsion data
\footnote{$^)$It is worth noting that the new recent emulsion KEK-PS E373 [20] and J-PARC E07 [21] experiments,
in which the remarkable events named, respectively, "KISO" and "IBUKI" were observed, reported the evidence
of a likely the Coulomb-assisted nuclear 1$p$ single-particle $\Xi^-$ state of the bound
$\Xi^-$--$^{14}$N(g.s.) system with binding energy of about 1 MeV, which also testifies in favor of shallow
$\Xi^-$--nucleus potential.}$^)$,
from missing-mass
measurements [22--25] in the inclusive ($K^-$,$K^+$) reaction on nuclear targets at incident momenta of
1.6--1.8 GeV/c in the $\Xi^-$ bound and quasi-free regions with insufficiently good (10 MeV or worse) [22--24]
and moderate (5.4 MeV) [25] energy resolutions, from the analysis the results of these measurements
yet in Refs. [26, 27] using Green's function method of the DWIA, the theoretical predictions [28--31] that the
$\Xi^-$ cascade hyperon feels only a weak attractive potential in nuclei, the depth of which is not so large,
$\sim$ -(10--20) MeV at central nuclear density and at rest as well as the
predictions [32, 33] within the SU(6) quark model that the $\Xi^-$ single-particle potential in symmetric
nuclear matter becomes almost zero at high $\Xi^-$ momenta around 1.3 GeV/c corresponding to the $K_1(1270)^+$
momentum range of $\sim$ 1--2 GeV/c where the $K_1(1270)^+$ production cross sections are the greatest, we
will ignore the modification of the free space mass $m_{\Xi^-}$ of the $\Xi^-$ hyperons,
produced together with the $K_1(1270)^+$ mesons in the process (1), in the nuclear medium.
To improve essentially our understanding of the $\Xi^-N$ interaction the high-quality data on the doubly
strange $S=-2$ hypernuclei are needed. It is expected that in the near future high resolution (up to 1.4 MeV)
and high statistics data on the missing-mass spectra for the $^{12}$C($K^-$,$K^+$) reaction around the $\Xi^-$
production threshold and the X-ray data on the level shifts and width broadening of the $\Xi^-$ atomic states in the
$\Xi^-$ atoms will become available from the planned dedicated E70 and E03 experiments at the J-PARC Hadron
Experimental Facility [34].
Accounting for that the in-medium threshold energy
\footnote{$^)$Determining mainly the strength of the $K_1(1270)^+$ production cross sections in
near-threshold antikaon--nucleus collisions.}$^)$
$\sqrt{s^*_{\rm th}}=<m_{K_1}^*>+m_{\Xi^-}$
of the process (1) looks like that for the final charged particles, influenced also by the respective Coulomb
potentials, due to the cancelation of these potentials,
we will neglect here their impact on these particles as well.

The total energy $E^\prime_{K_1}$ of the $K_1(1270)^+$ meson in nuclear matter is
expressed via its average effective mass $<m^*_{K_1}>$ and its in-medium momentum
${\bf p}^{\prime}_{K_1}$ by the expression [16]:
\begin{equation}
E^\prime_{K_1}=\sqrt{({\bf p}^{\prime}_{K_1})^2+(<m^*_{K_1}>)^2}.
\end{equation}
The momentum ${\bf p}^{\prime}_{K_1}$ is related to the vacuum $K_1(1270)^+$
momentum ${\bf p}_{K_1}$ as follows [16]:
\begin{equation}
E^\prime_{K_1}=\sqrt{({\bf p}^{\prime}_{K_1})^2+(<m^*_{K_1}>)^2}=
\sqrt{{\bf p}^2_{K_1}+m^2_{K_1}}=E_{K_1},
\end{equation}
where $E_{K_1}$ is the $K_1(1270)^+$ total energy in vacuum.

As the $K_1(1270)^+$--nucleon elastic cross section is expected to be small similar to the $K^*(892)^+N$
elastic cross section [35], we will ignore quasielastic ${K_1(1270)^+}N$ rescatterings in the present work.
Then, taking into account the attenuation of the incident antikaon and the final $K_1(1270)^+$ meson
in the nuclear matter in terms, respectively, of the $K^-N$ total cross section $\sigma_{{K^-}N}^{\rm tot}$
and the total width $\Gamma_{K_1}(|{\bf r}|)$ in the $K_1(1270)^+$ rest frame, taken at the point ${\bf r}$
inside the nucleus, as well as using the results given in [16], we represent the inclusive differential
cross section for the production of $K_1(1270)^+$ mesons with vacuum momentum
${\bf p}_{K_1}$ on nuclei in the direct process (1) as follows:
\begin{equation}
\frac{d\sigma_{{K^-}A \to {K_1(1270)^+}X}^{({\rm prim})}
({\bf p}_{K^-},{\bf p}_{K_1})}
{d{\bf p}_{K_1}}=I_{V}[A,\theta_{K_1}]
\left(\frac{Z}{A}\right)\left<\frac{d\sigma_{{K^-}p\to K_1(1270)^+{{\Xi^-}}}({\bf p}_{K^-},
{\bf p}^{\prime}_{{K_1}})}{d{\bf p}^{\prime}_{{K_1}}}\right>_A\frac{d{\bf p}^{\prime}_{{K_1}}}
{d{\bf p}_{{K_1}}},
\end{equation}
where
\begin{equation}
I_{V}[A,\theta_{{K_1}}]=A\int\limits_{0}^{R}r_{\bot}dr_{\bot}
\int\limits_{-\sqrt{R^2-r_{\bot}^2}}^{\sqrt{R^2-r_{\bot}^2}}dz
\rho(\sqrt{r_{\bot}^2+z^2})
\exp{\left[-\sigma_{{K^-}N}^{\rm tot}A\int\limits_{-\sqrt{R^2-r_{\bot}^2}}^{z}
\rho(\sqrt{r_{\bot}^2+x^2})dx\right]}
\end{equation}
$$
\times
\int\limits_{0}^{2\pi}d{\varphi}\exp{\left[-
\int\limits_{0}^{l(\theta_{{K_1}},\varphi)}\frac{dx}
{\lambda_{K_1}(\sqrt{x^2+2a(\theta_{{K_1}},\varphi)x+b+R^2})}\right]};
$$
\begin{equation}
a(\theta_{K_1},\varphi)=z\cos{\theta_{K_1}}+
r_{\bot}\sin{\theta_{K_1}}\cos{\varphi},\,\,\,\,b=r_{\bot}^2+z^2-R^2,
\end{equation}
\begin{equation}
l(\theta_{K_1},\varphi)=\sqrt{a^2(\theta_{K_1},\varphi)-b}-
a(\theta_{K_1},\varphi),
\end{equation}
\begin{equation}
\sigma_{{K^-}N}^{\rm tot}=\frac{Z}{A}\sigma_{{K^-}p}^{\rm tot}+\frac{N}{A}\sigma_{{K^-}n}^{\rm tot},
\end{equation}
\begin{equation}
\lambda_{K_1}(|{\bf r}|)=\frac{p^{\prime}_{K_1}}{m^*_{K_1}(|{\bf r}|)\Gamma_{K_1}(|{\bf r}|)},\,\,\,
m^*_{K_1}(|{\bf r}|)=m_{K_1}+V_0\frac{{\rho_N}(|{\bf r}|)}{{\rho_0}}
\end{equation}
and
\begin{equation}
\left<\frac{d\sigma_{{K^-}p\to K_1(1270)^+{\Xi^-}}({\bf p}_{K^-},{\bf p}^{\prime}_{K_1})}
{d{\bf p}^{\prime}_{K_1}}\right>_A=
\int\int
P_A({\bf p}_t,E)d{\bf p}_tdE
\end{equation}
$$
\times
\left\{\frac{d\sigma_{{K^-}p\to K_1(1270)^+{\Xi^-}}[\sqrt{s^*},<m_{K_1}^*>,
m_{\Xi^-},{\bf p}^{\prime}_{K_1}]}
{d{\bf p}^{\prime}_{K_1}}\right\},
$$
\begin{equation}
  s^*=(E_{K^-}+E_t)^2-({\bf p}_{K^-}+{\bf p}_t)^2,
\end{equation}
\begin{equation}
   E_t=M_A-\sqrt{(-{\bf p}_t)^2+(M_{A}-m_{N}+E)^{2}}.
\end{equation}
Here,
$d\sigma_{{K^-}p\to {K_1(1270)^+}{\Xi^-}}[\sqrt{s^*},<m_{K_1}^*>,m_{\Xi^-},{\bf p}^{\prime}_{K_1}]
/d{\bf p}^{\prime}_{K_1}$
is the off-shell inclusive differential cross section for the production of ${K_1(1270)^+}$ meson and
$\Xi^-$ hyperon with modified mass $<m_{K_1}^*>$ and free mass $m_{\Xi^-}$, respectively.
The $K_1(1270)^+$ meson is produced with in-medium momentum
${\bf p}^{\prime}_{{K_1}}$ in process (1) at the ${K^-}p$ center-of-mass energy $\sqrt{s^*}$.
$E_{K^-}$ and ${\bf p}_{K^-}$ are the total energy and momentum of the incident antikaon
($E_{K^-}=\sqrt{m^2_{K}+{\bf p}^2_{K^-}}$, $m_{K}$ is the free space kaon mass);
$\rho({\bf r})$ and $P_A({\bf p}_t,E)$ are the local nucleon density and the
spectral function of the target nucleus A normalized to unity
(the concrete information about these quantities, used in the subsequent calculations, is given
in Refs. [36--39]);
${\bf p}_t$ and $E$ are the internal momentum and removal energy of the target proton
involved in the collision process (1);
$Z$ and $N$ are the numbers of protons and neutrons in the target nucleus ($A=Z+N$),
$M_{A}$  and $R$ are its mass and radius; $m_N$ is the free space nucleon mass;
$\theta_{K_1}$ is the polar angle of
vacuum momentum ${\bf p}_{{K_1}}$ in the laboratory system with z-axis directed along the momentum
${\bf p}_{{K^-}}$ of the incident antikaon beam;
$\sigma_{{K^-}p}^{\rm tot}$ and $\sigma_{{K^-}n}^{\rm tot}$
are the total cross sections of the free ${K^-}p$ and ${K^-}n$ interactions at beam momenta belonging
to the range of 2.5--3.5 GeV/c
\footnote{$^)$ We ignore the medium modification of the incoming $K^-$ meson mass in this momentum
range [40].}$^)$
.
In this range $\sigma_{{K^-}p}^{\rm tot}\approx$ 28 mb and
$\sigma_{{K^-}n}^{\rm tot}\approx$ 22 mb [41]. With these, the quantity $\sigma_{{K^-}N}^{\rm tot}$,
entering into Eqs. (6) and (9), amounts approximately to 25 mb for both considered target nuclei
$^{12}$C and $^{184}$W. We will use this value in our present calculations.

   According to [16], we suppose that the off-shell differential cross section
$d\sigma_{{K^-}p \to {K_1(1270)^+}{\Xi^-}}[\sqrt{s^*},<m_{K_1}^*>,m_{\Xi^-},{\bf p}^{\prime}_{K_1}]
/d{\bf p}^{\prime}_{K_1}$ for $K_1(1270)^+$ production in process (1) is equivalent to
the respective on-shell cross section calculated for the off-shell kinematics of this process
as well as for the final $K_1(1270)^+$ meson and $\Xi^-$ hyperon in-medium mass $<m_{K_1}^*>$
and free mass $m_{\Xi^-}$, respectively. Taking into account the two-body kinematics of the process (1),
we get the following expression for the differential cross section
$d\sigma_{{K^-}p \to {K_1(1270)^+}{\Xi^-}}[\sqrt{s^*},<m_{K_1}^*>,m_{\Xi^-},{\bf p}^{\prime}_{K_1}]
/d{\bf p}^{\prime}_{K_1}$:
\begin{equation}
\frac{d\sigma_{K^{-}p \to {K_1(1270)^+}{\Xi^-}}[\sqrt{s^*},<m_{K_1}^*>,m_{\Xi^-},{\bf p}^{\prime}_{K_1}]}
{d{\bf p}^{\prime}_{K_1}}=
\frac{\pi}{I_2[s^*,<m_{K_1}^*>,m_{\Xi^-}]E^{\prime}_{K_1}}
\end{equation}
$$
\times
\frac{d\sigma_{{K^{-}}p \to {K_1(1270)^+}{\Xi^-}}(\sqrt{s^*},<m_{K_1}^*>,m_{\Xi^-},\theta^*_{K_1})}
{d{\bf \Omega}^*_{K_1}}
$$
$$
\times
\frac{1}{(\omega+E_t)}\delta\left[\omega+E_t-\sqrt{m_{\Xi^-}^2+({\bf Q}+{\bf p}_t)^2}\right],
$$
where
\begin{equation}
I_2[s^*,<m_{K_1}^*>,m_{\Xi^-}]=\frac{\pi}{2}
\frac{\lambda[s^*,(<m_{K_1}^*>)^{2},m_{\Xi^-}^{2}]}{s^*},
\end{equation}
\begin{equation}
\lambda(x,y,z)=\sqrt{{\left[x-({\sqrt{y}}+{\sqrt{z}})^2\right]}{\left[x-
({\sqrt{y}}-{\sqrt{z}})^2\right]}},
\end{equation}
\begin{equation}
\omega=E_{K^-}-E^{\prime}_{K_1}, \,\,\,\,{\bf Q}={\bf p}_{K^-}-{\bf p}^{\prime}_{K_1}.
\end{equation}
Here,
$d\sigma_{{K^{-}}p \to {K_1(1270)^+}{\Xi^-}}(\sqrt{s^*},<m_{K_1}^*>,m_{\Xi^-},\theta^*_{K_1})
/d{\bf \Omega}^*_{K_1}$
is the off-shell differential cross section for the production of $K_1(1270)^+$ mesons
in process (1) under the polar angle $\theta^*_{K_1}$ in the ${K^-}p$ c.m.s. This cross section
is assumed to be isotropic in our calculations of $K_1(1270)^+$ meson production in ${K^-}A$ interactions:
\begin{equation}
\frac{d\sigma_{{K^{-}}p \to {K_1(1270)^+}{\Xi^-}}(\sqrt{s^*},<m_{K_1}^*>,m_{\Xi^-},\theta^*_{K_1})}
{d{\bf \Omega}^*_{K_1}}=\frac{\sigma_{{K^{-}}p \to {K_1(1270)^+}{\Xi^-}}(\sqrt{s^*},\sqrt{s^*_{\rm th}})}{4\pi}.
\end{equation}
Here, $\sigma_{{K^{-}}p \to {K_1(1270)^+}{\Xi^-}}(\sqrt{s^*},\sqrt{s^*_{\rm th}})$ is the
"in-medium" total cross section of reaction (1) having the threshold energy $\sqrt{s^*_{\rm th}}$ defined above.
In line with above-mentioned, it is equivalent to the vacuum cross section
$\sigma_{{K^{-}}p \to {K_1(1270)^+}{\Xi^-}}(\sqrt{s},\sqrt{s_{\rm th}})$, in which the vacuum threshold energy
$\sqrt{s_{\rm th}}=m_{K_1}+m_{\Xi^-}=2.594$ GeV is replaced by the in-medium one $\sqrt{s^*_{\rm th}}$
and the free center-of-mass energy squared $s$, presented by the formula
\begin{equation}
s=(E_{K^-}+m_N)^2-{\bf p}_{K^-}^2,
\end{equation}
is replaced by the in-medium one $s^*$, defined by the expression (12).

For the free total cross section
$\sigma_{{K^{-}}p \to {K_1(1270)^+}{\Xi^-}}(\sqrt{s},\sqrt{s_{\rm th}})$ we have adopted the
following estimates, based on the available only two experimental data points (6.2$\pm$0.6) ${\rm \mu}$b and
(1.7$\pm$0.5) ${\rm \mu}$b for the vacuum total cross sections of the processes
$K^-p \to K_1(1270)^+\Xi^- \to K{\rho}\Xi^-$ and $K^-p \to K_1(1270)^+\Xi^- \to K^*(892){\pi}\Xi^-$
at 0.4 GeV excess energy $\sqrt{s}-\sqrt{s_{\rm th}}$ (or at the beam momentum of 4.15 GeV/c) [42]
as well as on the relatively large amount of existing experimental data for the total cross section of the
$K^-p \to K^+\Xi^-$ reaction [41]. Accounting for the branching ratios of (42$\pm$6)\% and (16$\pm$5)\% for the
channels $K_1(1270)^+ \to K{\rho}$ and $K_1(1270)^+ \to K^*(892){\pi}$ and these data points, the values of
(14.8$\pm$2.6) ${\rm \mu}$b and (10.6$\pm$4.6) ${\rm \mu}$b for the free cross section
$\sigma_{{K^{-}}p \to {K_1(1270)^+}{\Xi^-}}(\sqrt{s},\sqrt{s_{\rm th}})$ are obtained at
$\sqrt{s}-\sqrt{s_{\rm th}}=$0.4 GeV. They are shown in Fig. 1 by full triangles.
Here, the vacuum total cross section of the reaction $K^-p \to K^+\Xi^-$ is presented as well as a function
of the respective excess energy $\sqrt{{\tilde s}}-\sqrt{{\tilde s}_{\rm th}}$ available in this reaction
with threshold energy $\sqrt{{\tilde s}_{\rm th}}=m_K+m_{\Xi^-}$.
Full squares are experimental data taken from the compilation of Flaminio {\it et al.} [41].
Solid curve is their parametrization by the formula
\begin{equation}
\sigma_{{K}^-p \to {K^+}{\Xi^-}}(\sqrt{{\tilde s}},\sqrt{{\tilde s}_{\rm th}})=
	235.6\left(1-\sqrt{{\tilde s}_{\rm th}}/\sqrt{{\tilde s}}\right)^{2.4}
               \left(\sqrt{{\tilde s}_{\rm th}}/\sqrt{{\tilde s}}\right)^{16.6}~[{\rm mb}],
\end{equation}
suggested in Ref. [43].
\begin{figure}[htb]
\begin{center}
\includegraphics[width=18.0cm]{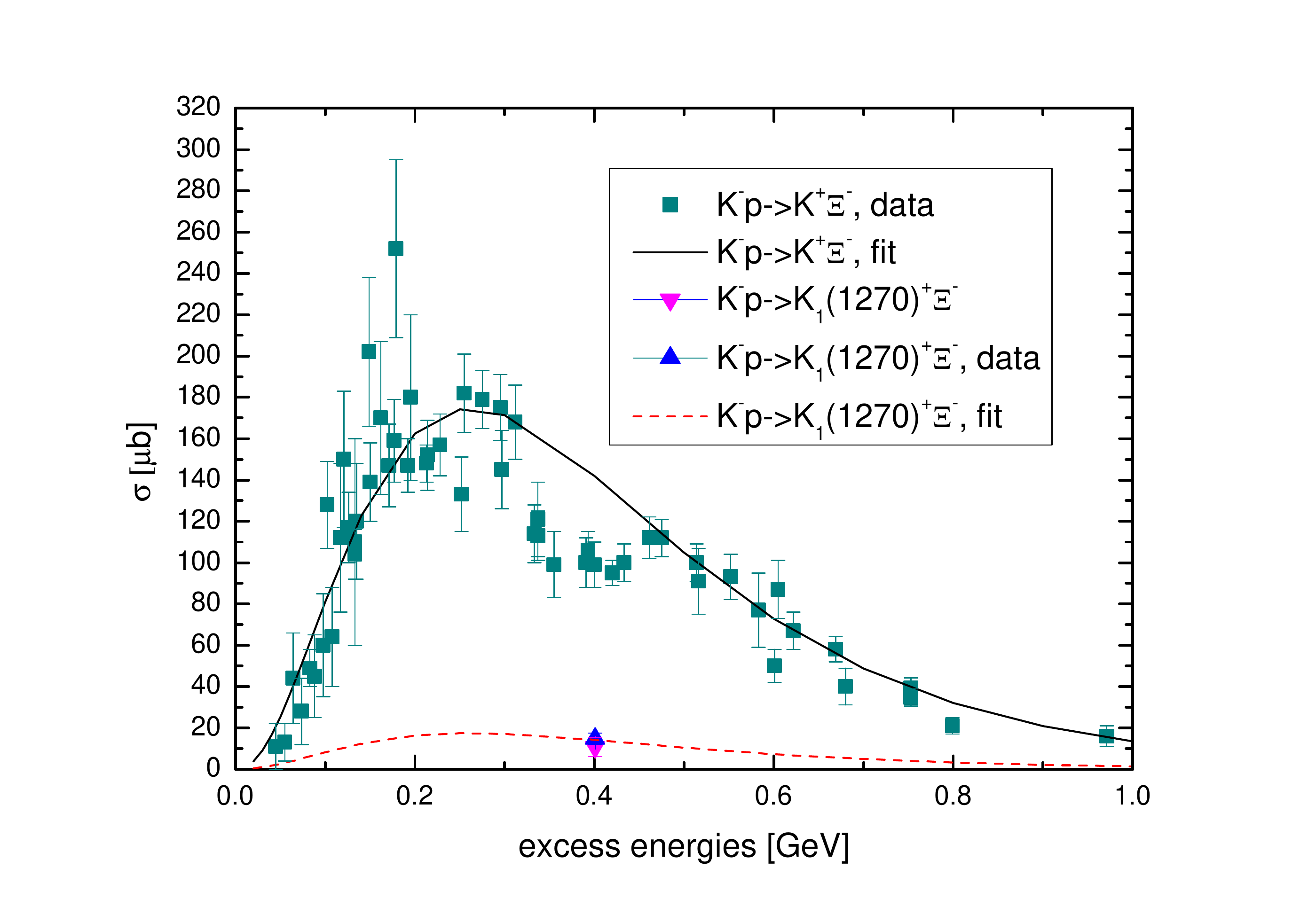}
\vspace*{-2mm} \caption{(Color online) Total cross sections for the reactions $K^-p \to {K^+}{\Xi^-}$
and $K^-p \to {K_1(1270)^+}{\Xi^-}$ as functions of the available excess energies
$\sqrt{{\tilde s}}-\sqrt{{\tilde s}_{\rm th}}$ and $\sqrt{s}-\sqrt{s_{\rm th}}$
above their thresholds $\sqrt{{\tilde s}_{\rm th}}$ and $\sqrt{s_{\rm th}}$, respectively.
For notation see the text.}
\label{void}
\end{center}
\end{figure}
\begin{figure}[!h]
\begin{center}
\includegraphics[width=18.0cm]{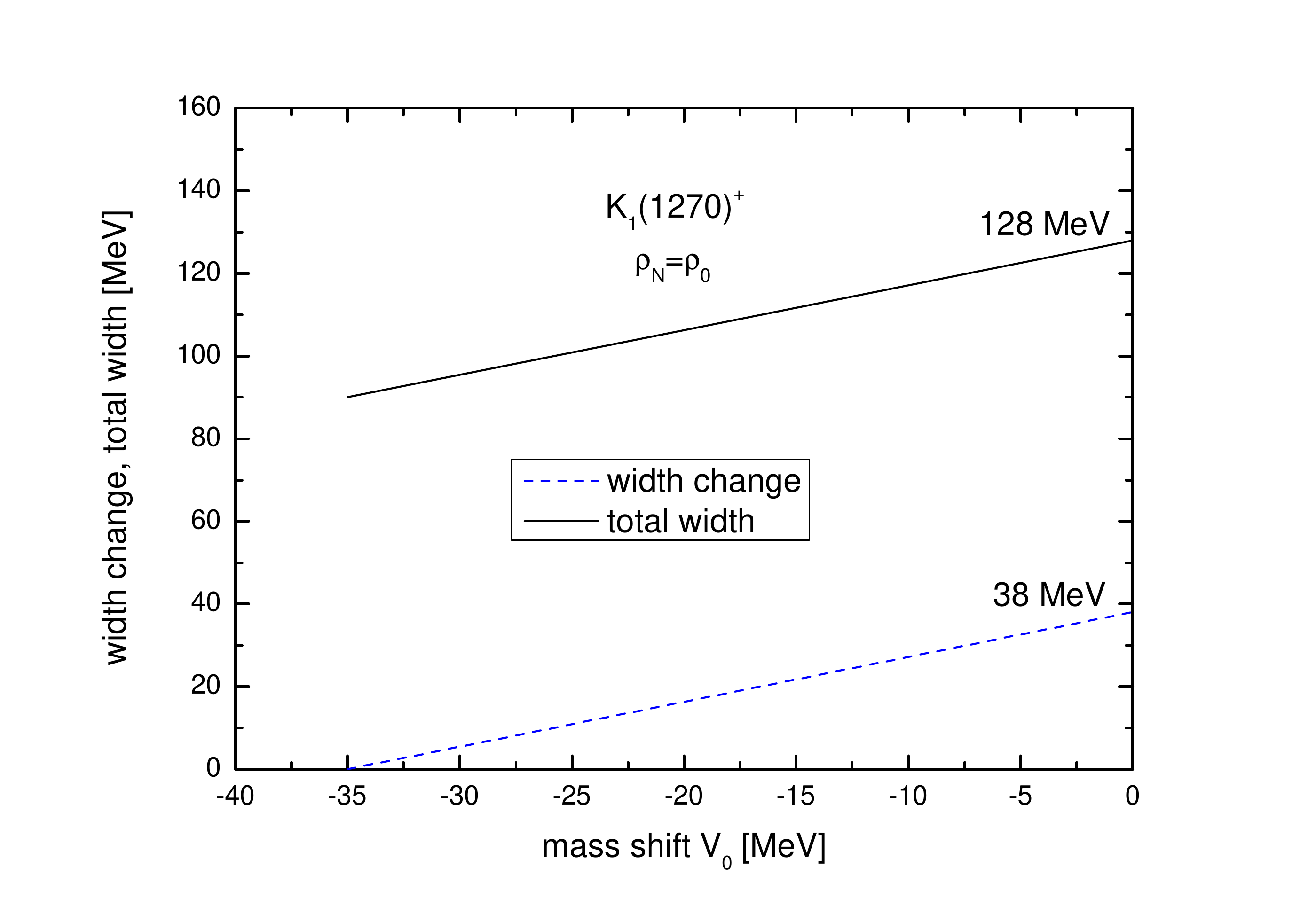}
\vspace*{-2mm} \caption{(Color online) In-medium change of the $K_1(1270)^+$ meson width
and its total in-medium width in its rest frame
as functions of the $K_1(1270)^+$ in-medium mass shift $V_0$ at normal nuclear matter density.}
\label{void}
\end{center}
\end{figure}
Visual inspection of Fig. 1 yields the following ratio of $K_1(1270)^+{\Xi^-}$ to $K^+{\Xi^-}$
production cross sections at the same excess energies $\sqrt{s}-\sqrt{s_{\rm th}}$ and
$\sqrt{{\tilde s}}-\sqrt{{\tilde s}_{\rm th}}$ above the $K_1(1270)^+{\Xi^-}$ and $K^+{\Xi^-}$
thresholds equal to 0.4 GeV:
\begin{equation}
\sigma_{{K^{-}}p \to {K_1(1270)^+}{\Xi^-}}(\sqrt{s},\sqrt{s_{\rm th}})/
\sigma_{{K}^-p \to {K^+}{\Xi^-}}(\sqrt{{\tilde s}},\sqrt{{\tilde s}_{\rm th}})\approx 0.1.
\end{equation}
Since no data on the total cross section
$\sigma_{{K^{-}}p \to {K_1(1270)^+}{\Xi^-}}(\sqrt{s},\sqrt{s_{\rm th}})$ exist for other excess
energies, it is natural to estimate it at them in the threshold region assuming that the ratio (21)
is also valid at the same excess energies $\sqrt{s}-\sqrt{s_{\rm th}}$ and
$\sqrt{{\tilde s}}-\sqrt{{\tilde s}_{\rm th}}$ differing from 0.4 GeV. Under this assumption, the
center-of-mass energies $\sqrt{{\tilde s}}$ and $\sqrt{s}$, at which the cross sections entering
into the ratio (21) are calculated, are linked by the relation
\begin{equation}
\sqrt{{\tilde s}}-\sqrt{{\tilde s}_{\rm th}}=\sqrt{s}-\sqrt{s_{\rm th}}.
\end{equation}
Thus, we have
\begin{equation}
\sqrt{{\tilde s}}=\sqrt{s}-\sqrt{s_{\rm th}}+\sqrt{{\tilde s}_{\rm th}}=\sqrt{s}-m_{K_1}+m_{K}.
\end{equation}
When reactions ${K}^-p \to {K^+}{\Xi^-}$ and ${K^{-}}p \to {K_1(1270)^+}{\Xi^-}$ go in medium on
an off-shell target protons, instead of the quantity $\sqrt{{\tilde s}_{\rm th}}$,
appearing in Eqs. (20)--(23), we should use
the in-medium threshold energy $\sqrt{{\tilde s}^*_{\rm th}}=<m_{K}^*>+m_{\Xi^-}$, where
$<m_{K}^*>$ is the $K^+$ meson average in-medium mass defined by the expression analogous to that given
by Eq. (2) and in which $V_0=+22$ MeV [38].
And instead of the quantities $\sqrt{s}$ and $\sqrt{s_{\rm th}}$, entering into Eqs. (21)--(23),
one needs to adopt, respectively, the in-medium expression (12) and in-medium threshold energy
$\sqrt{ s^*_{\rm th}}$ determined above. The foregoing leads, as is easy to see, to the following substitutions
$m_{K_1} \to$ $<m_{K_1}^*>$, $m_{K} \to$ $<m_{K}^*>$ in the second relation of Eq. (23).

At incident antikaon momenta of interest $p_{K^-}$ $\le$ 3.5 GeV/c, $\sqrt{s} \le 2.785$ GeV and
$\sqrt{{\tilde s}} \le 2.007$ GeV. Taking into account that the relation between the free collision
energy squared ${\tilde s}$ and the total energy ${\tilde E}_{K^-}$ and momentum ${\tilde p}_{K^-}$
of the $K^-$ meson inducing the reaction ${K^{-}}p \to {K^+}{\Xi^-}$ is similar to that given above
by Eq. (19), we easily get that the latter
corresponds to its  laboratory momenta ${\tilde p}_{K^-}$ $\le$ 1.466 GeV/c. They are not far
from the threshold momentum of 1.047 GeV/c of this reaction. For the free total
cross section $\sigma_{{K}^-p \to {K^+}{\Xi^-}}(\sqrt{{\tilde s}},\sqrt{{\tilde s}_{\rm th}})$
at these momenta we will employ in our calculations the expression (20). Using it and the relations
(21)--(23), we calculated the vacuum total cross section
$\sigma_{{K^{-}}p \to {K_1(1270)^+}{\Xi^-}}(\sqrt{s},\sqrt{s_{\rm th}})$ of the process (1).
It is plotted with a dashed curve in Fig. 1.
As can be seen from this figure, the on-shell cross section $\sigma_{{K}^-p \to {K_1(1270)^+}{\Xi^-}}$
amounts approximately to 16 ${\rm \mu}$b for the initial antikaon momentum of 3.5 GeV/c, corresponding
to the excess energy $\sqrt{s}-\sqrt{s_{\rm th}}$ of about 0.2 GeV.
This offers the possibility of measuring the $K_1(1270)^+$ yield in $K^-A$ collisions both at the
above-threshold and below-threshold beam momenta at the J-PARC Hadron Experimental Facility with
sizable strength. Here, the $K_1(1270)^+$ mesons could be registered via the hadronic decays
$K_1(1270)^+ \to K{\rho}$ and $K_1(1270)^+ \to K^*(892){\pi}$.
\begin{figure}[!h]
\begin{center}
\includegraphics[width=18.0cm]{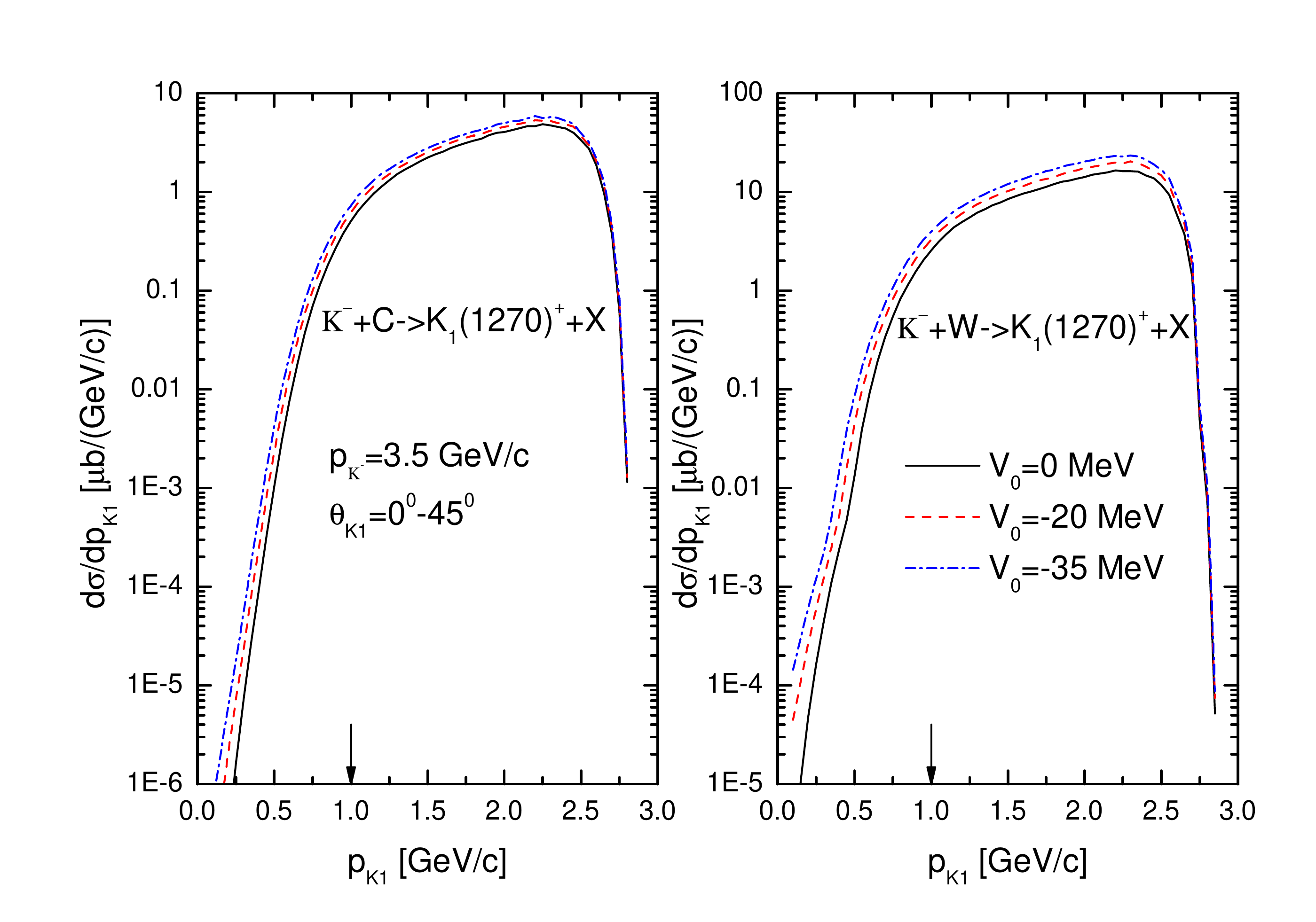}
\vspace*{-2mm} \caption{(Color online) Momentum differential cross sections for the production of $K_1(1270)^+$
mesons from the direct ${K^-}p \to {K_1(1270)^+}{\Xi^-}$ channel in the laboratory polar angular range of
0$^{\circ}$--45$^{\circ}$ in the interaction of $K^-$ mesons of momentum of 3.5 GeV/c with $^{12}$C
(left) and $^{184}$W (right) nuclei, calculated for different values of the $K_1(1270)^+$ meson effective
scalar potential $V_0$ at density $\rho_0$ indicated in the inset. The arrows indicate the boundary between
the low-momentum and high-momentum regions of the $K_1(1270)^+$ spectra.}
\label{void}
\end{center}
\end{figure}
\begin{figure}[!h]
\begin{center}
\includegraphics[width=18.0cm]{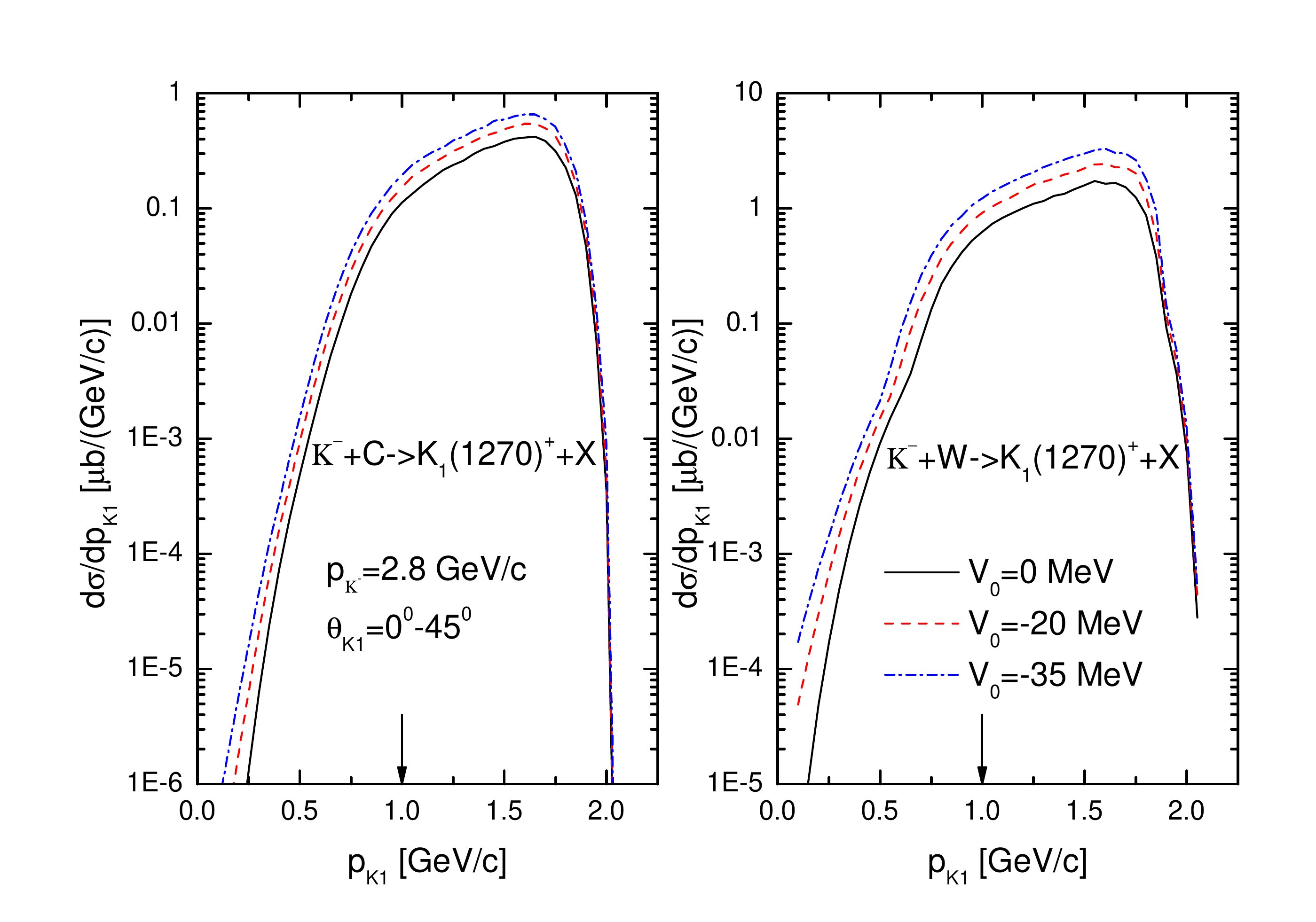}
\vspace*{-2mm} \caption{(Color online) The same as in Fig. 3, but for the incident antikaon beam momentum of
2.8 GeV/c.}
\label{void}
\end{center}
\end{figure}
\begin{figure}[!h]
\begin{center}
\includegraphics[width=18.0cm]{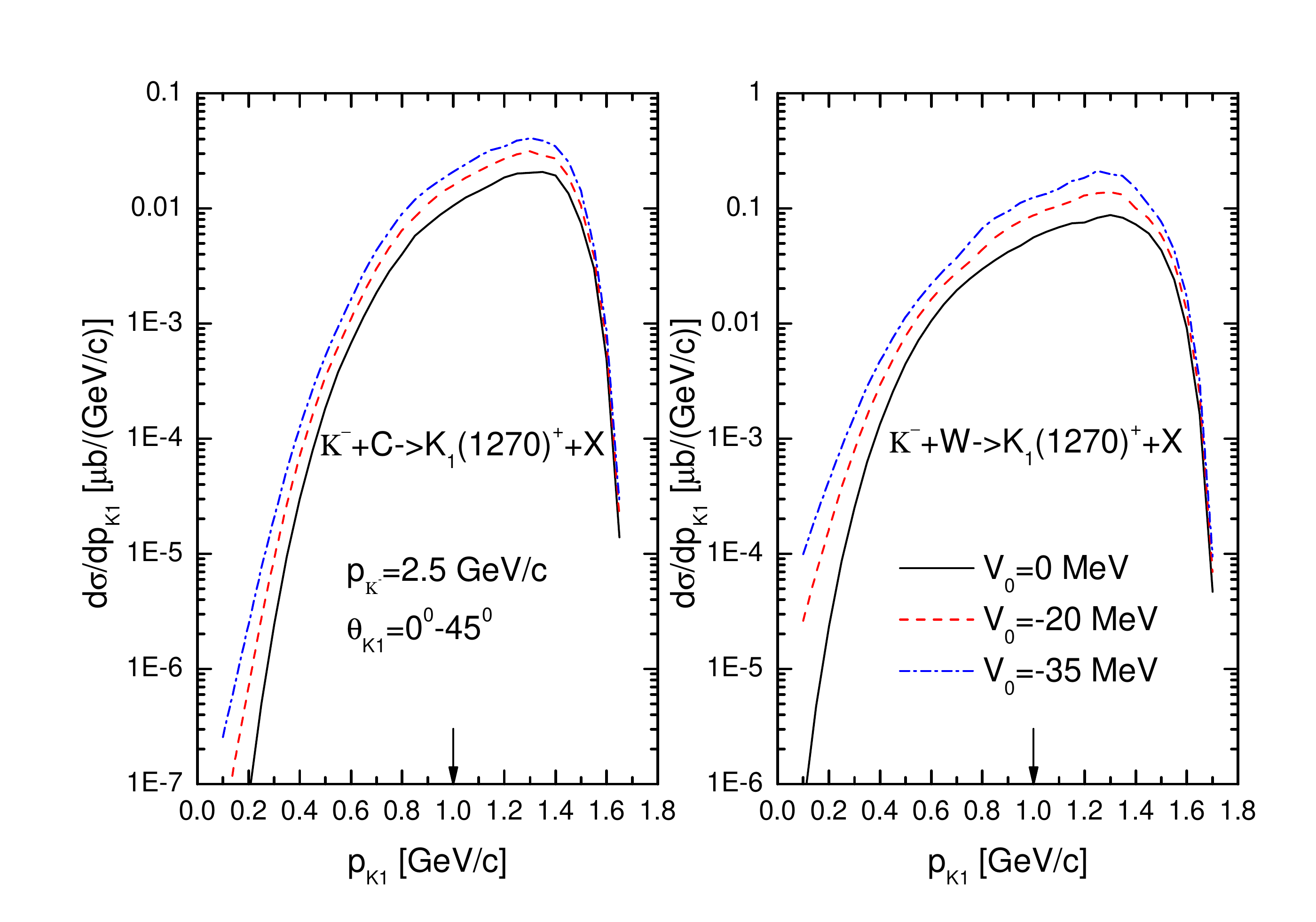}
\vspace*{-2mm} \caption{(Color online) The same as in Fig. 3, but for the incident antikaon beam momentum of
2.5 GeV/c.}
\label{void}
\end{center}
\end{figure}
\begin{figure}[htb]
\begin{center}
\includegraphics[width=18.0cm]{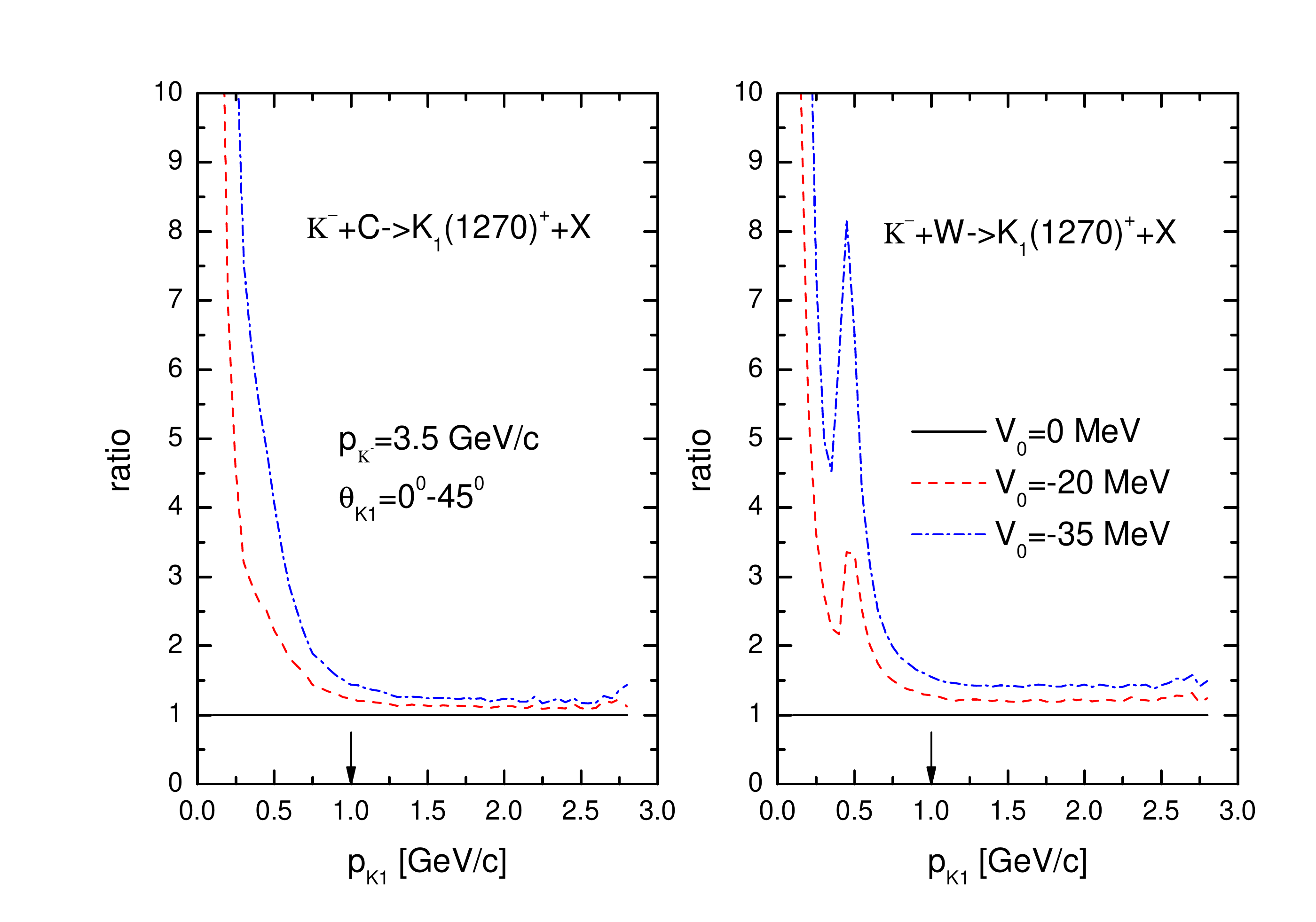}
\vspace*{-2mm} \caption{(Color online) Ratio between the differential cross sections for $K_1(1270)^+$
production on $^{12}$C (left) and $^{184}$W (right) target nuclei in the angular range of
0$^{\circ}$--45$^{\circ}$ by 3.5 GeV/c $K^-$ mesons,
calculated with and without the $K_1(1270)^+$ meson in-medium mass shift
$V_0$ at density $\rho_0$ indicated in the inset, as a function of $K_1(1270)^+$ momentum.
The arrows indicate the boundary between
the low-momentum and high-momentum regions of the $K_1(1270)^+$ spectra.}
\label{void}
\end{center}
\end{figure}
\begin{figure}[!h]
\begin{center}
\includegraphics[width=18.0cm]{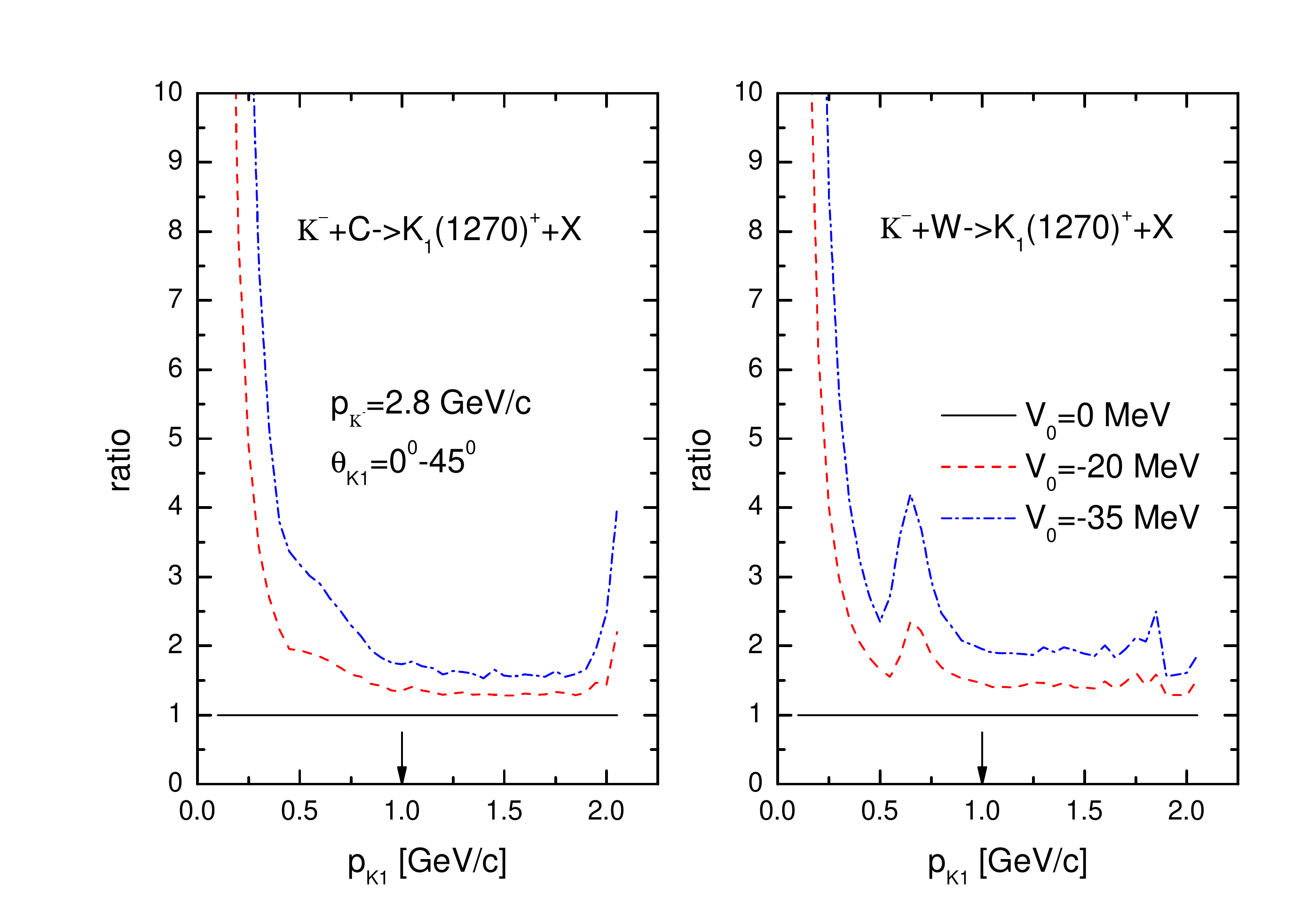}
\vspace*{-2mm} \caption{(Color online) The same as in Fig. 6, but for the incident antikaon beam momentum of
2.8 GeV/c.}
\label{void}
\end{center}
\end{figure}
\begin{figure}[!h]
\begin{center}
\includegraphics[width=18.0cm]{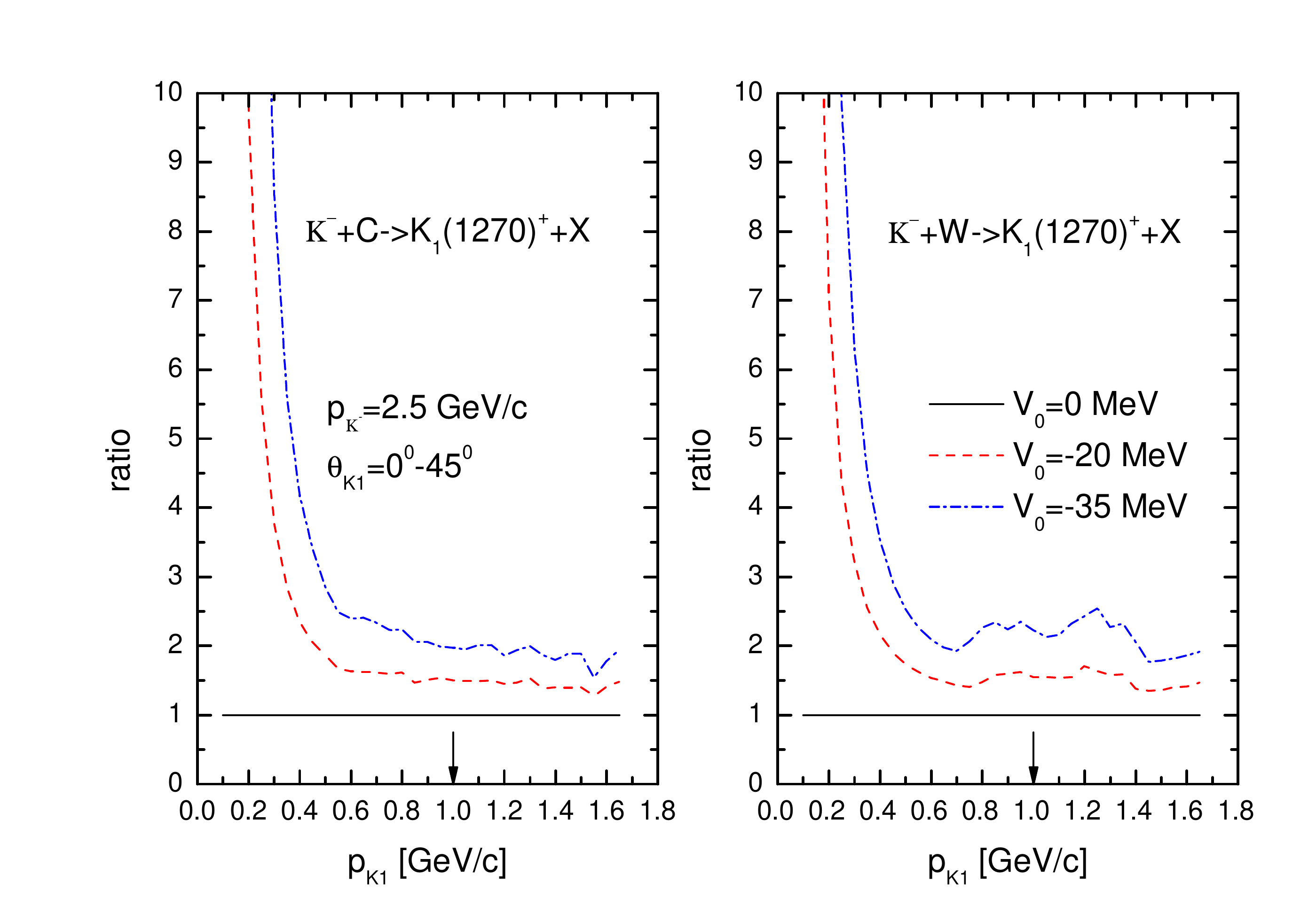}
\vspace*{-2mm} \caption{(Color online) The same as in Fig. 6, but for the incident antikaon beam momentum of
2.5 GeV/c.}
\label{void}
\end{center}
\end{figure}
\begin{figure}[!h]
\begin{center}
\includegraphics[width=18.0cm]{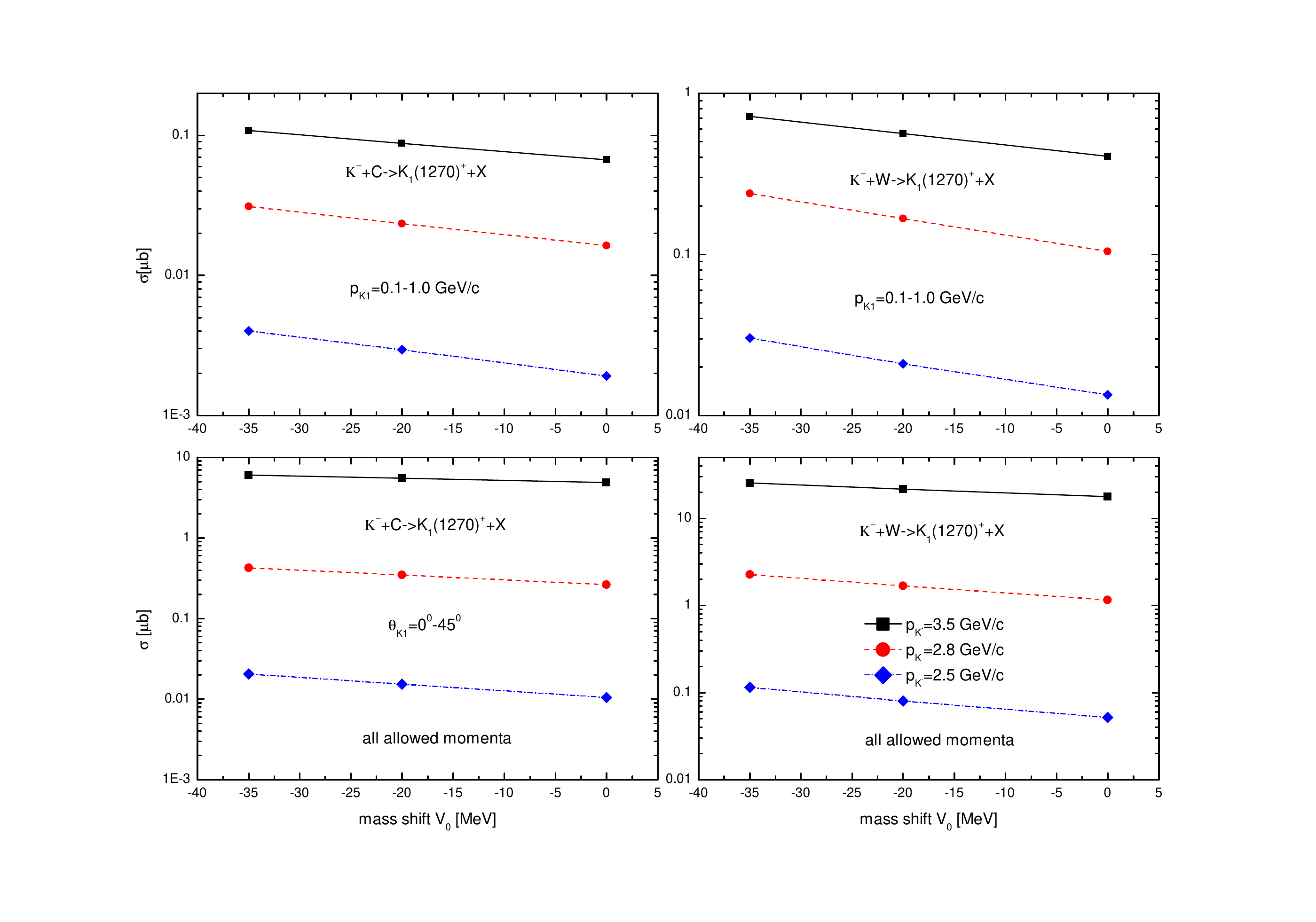}
\vspace*{-2mm} \caption{(Color online) Total cross sections for the production of $K_1(1270)^+$
mesons from the direct ${K^-}p \to {K_1(1270)^+}{\Xi^-}$ channel on C and W target nuclei with
momenta of 0.1--1.0 GeV/c (upper two panels) and with all allowed momenta $\ge$ 0.1 GeV/c at given beam momentum
(lower two panels) in the laboratory polar angular range of 0$^{\circ}$--45$^{\circ}$
by 2.5, 2.8 and 3.5 GeV/c $K^-$ mesons as functions of the $K_1(1270)^+$
in-medium mass shift $V_0$ at normal nuclear density. The lines are visual guides.}
\label{void}
\end{center}
\end{figure}
\begin{figure}[!h]
\begin{center}
\includegraphics[width=18.0cm]{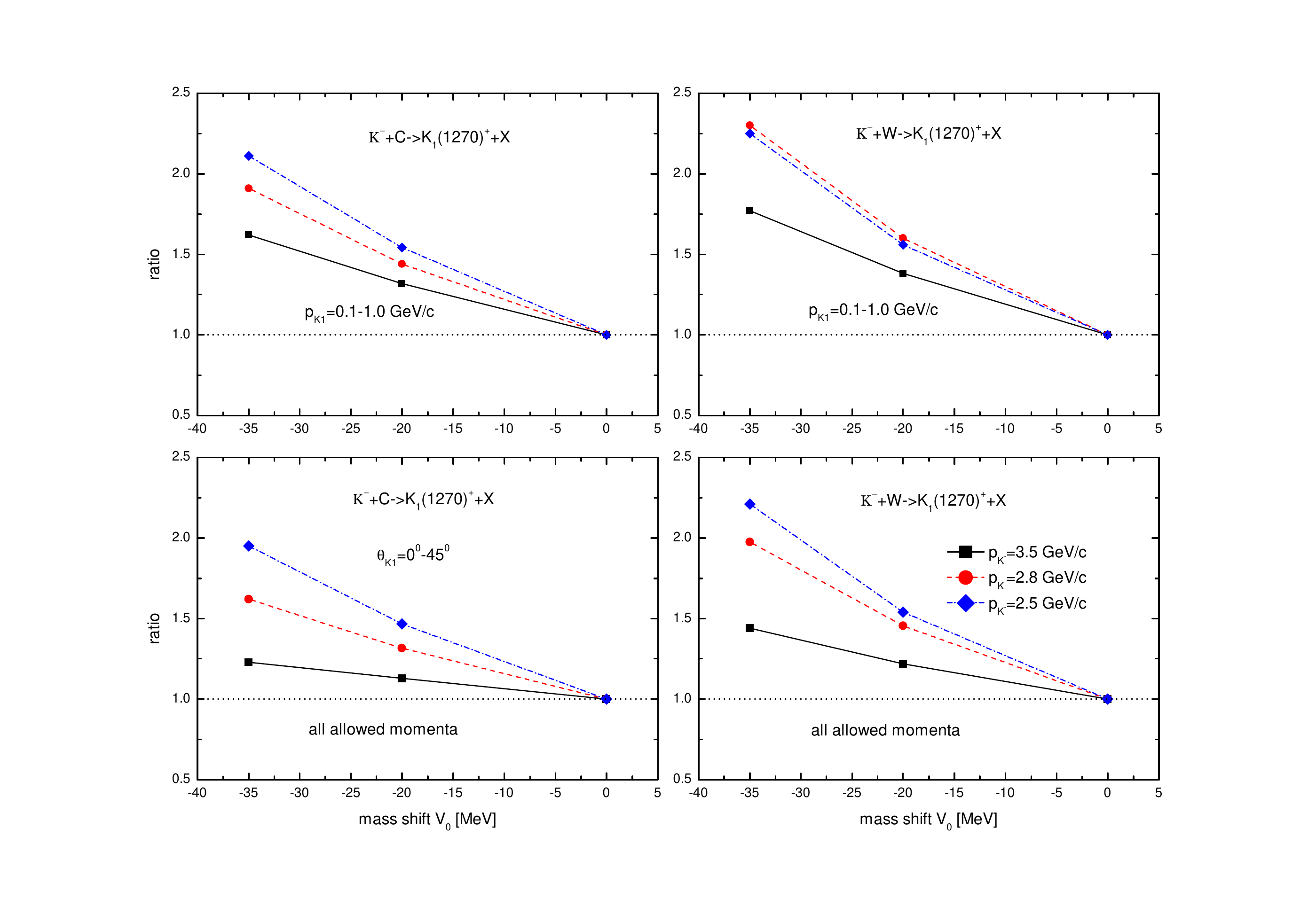}
\vspace*{-2mm} \caption{(Color online) Ratio between the total cross sections for the production of $K_1(1270)^+$
mesons from the direct ${K^-}p \to {K_1(1270)^+}{\Xi^-}$ channel
on $^{12}$C and $^{184}$W target nuclei at laboratory angles of 0$^{\circ}$--45$^{\circ}$ with
momenta of 0.1--1.0 GeV/c (upper two panels) and with all allowed momenta $\ge$ 0.1 GeV/c at given beam momentum
(lower two panels) by 2.5, 2.8 and 3.5 GeV/c $K^-$ mesons, calculated with and without the $K_1(1270)^+$
in-medium mass shift $V_0$ at normal nuclear density, as function of this shift. The lines are visual guides.}
\label{void}
\end{center}
\end{figure}

In Eqs. (6)--(8) we suppose that the direction of the $K_1(1270)^+$ meson three-momentum is not changed
during its propagation inside the nucleus  in the relatively weak nuclear
field, considered in the present work, from the production point here to the vacuum. As a result, the
quantities
$\left<d\sigma_{{K^-}p\to K_1(1270)^+{{\Xi^-}}}({\bf p}_{K^-},
{\bf p}^{\prime}_{{K_1}})/d{\bf p}^{\prime}_{{K_1}}\right>_A$ and
$d{\bf p}^{\prime}_{{K_1}}/d{\bf p}_{{K_1}}$, entering into Eq. (5), can be expressed in the
following simple forms
$\left<d\sigma_{{K^-}p\to K_1(1270)^+{{\Xi^-}}}(p_{K^-},
p^{\prime}_{{K_1}}, \theta_{K_1})/p^{\prime2}_{K_1}dp^{\prime}_{{K_1}}d{\bf \Omega}_{K_1}\right>_A$ and
$p^{\prime}_{{K_1}}/p_{{K_1}}$, where
${\bf \Omega}_{K_1}(\theta_{K_1},\varphi_{K_1})={\bf p}_{{K_1}}/p_{{K_1}}$.
Here, $\varphi_{K_1}$ is the azimuthal angle of the $K_1(1270)^+$ momentum ${\bf p}_{K_1}$
in the laboratory system.
Taking into account the fact that in the considered incident momentum
region $K_1(1270)^+$ mesons are mainly ejected, due to the kinematics, in a narrow cone along the beam line
\footnote{$^)$ Thus, for instance, at a beam momentum of 3.5 GeV/c the $K_1(1270)^+$ laboratory production
polar angles in reaction (1) proceeding on the target proton being at rest are $\le$ 18.5$^{\circ}$.}$^)$
,
we will calculate the $K_1(1270)^+$ momentum
differential and total production cross sections on $^{12}$C and $^{184}$W targets
for laboratory solid angle
${\Delta}{\bf \Omega}_{K_1}$=$0^{\circ} \le \theta_{K_1} \le 45^{\circ}$,
and $0 \le \varphi_{K_1} \le 2{\pi}$.
Integrating the full inclusive differential cross section (5) over this angular domain,
we can represent the differential cross section for $K_1(1270)^+$ meson
production in ${K^-}A$ reactions from the direct process (1), corresponding to this angle, in the following form:
\begin{equation}
\frac{d\sigma_{{K^-}A\to {K_1(1270)^+}X}^{({\rm prim})}
(p_{K^-},p_{K_1})}{dp_{K_1}}=
\int\limits_{{\Delta}{\bf \Omega}_{K_1}}^{}d{\bf \Omega}_{K_1}
\frac{d\sigma_{{K^-}A\to {K_1(1270)^+}X}^{({\rm prim})}
({\bf p}_{K^-},{\bf p}_{K_1})}{d{\bf p}_{K_1}}p_{K_1}^2
\end{equation}
$$
=2{\pi}\left(\frac{Z}{A}\right)\left(\frac{p_{K_1}}{p^{\prime}_{K_1}}\right)
\int\limits_{\cos45^{\circ}}^{1}d\cos{{\theta_{K_1}}}I_{V}[A,\theta_{K_1}]
\left<\frac{d\sigma_{{K^-}p\to {K_1(1270)^+}{\Xi^-}}(p_{K^-},
p^{\prime}_{K_1},\theta_{K_1})}{dp^{\prime}_{K_1}d{\bf \Omega}_{K_1}}\right>_A.
$$

We define now the $K_1(1270)^+$ meson total in-medium width $\Gamma_{K_1}(|{\bf r}|)$
appearing in Eq. (10) and adopted in our calculations of $K_1(1270)^+$ production in $K^-A$ interactions.
According to Ref. [8], we can represent it in the following form:
\begin{equation}
\Gamma_{K_1}(|{\bf r}|)=\Gamma_{K_1}+\Gamma_0\frac{{\rho_N}(|{\bf r}|)}{{\rho_0}},
\end{equation}
where $\Gamma_{K_1}=90$ MeV is the vacuum total $K_1(1270)^+$ decay width in its rest frame
and $\Gamma_0$ is the in-medium change of the $K_1(1270)^+$ width at saturation density $\rho_0$.
A work [8] finds a linear correlation between the width change $\Gamma_0$ and the in-medium mass shift $V_0$,
shown by dashed curve in Fig. 2. Analytically, a relation between the quantities $\Gamma_0$ and $V_0$
can be easily expressed as follows:
\begin{equation}
\Gamma_0=38~{\rm MeV}+(38~{\rm MeV}/35~{\rm MeV}){\cdot}V_0,\,\,\,  -35~{\rm MeV}\le V_0\le0~{\rm MeV}.
\end{equation}
One can see that the maximum change of the width is +38 MeV when the $K_1(1270)^+$ meson mass remains the same
in the medium, while for the upper limit of its mass shift of -35 MeV the width is not changed here at all.
In these cases, the resulting total in-medium width (25) of the $K_1(1270)^+$ meson,
depicted by the solid curve in Fig. 2, reaches the values of 128 and 90 MeV, respectively.
So, once the change of the $K_1(1270)^+$ meson
mass is allowed, its total in-medium width gets smaller. This will lead to a more larger $K_1(1270)^+$
decay mean "free" path $\lambda_{K_1}$, which in turn will cause more weaker absorption of
$K_1(1270)^+$ mesons in the nuclear matter.
Thus, for example, Eq. (10) shows that for typical values $p^{\prime}_{K_1} \approx m^*_{K_1}$ and total in-medium
$K_1(1270)^+$ meson decay width in its rest frame of 128 MeV, corresponding to no in-medium effects on its mass,
this mean "free" path is equal to 1.5 fm. Whereas in the case of maximum change of $K_1(1270)^+$ mass in the
medium, when the above total in-medium width equals to the vacuum width $\Gamma_{K_1}=90$ MeV,
the $K_1(1270)^+$ decay mean "free" path $\lambda_{K_1}$
is larger and is equal to 2.2 fm.
These values are comparable with the radius of $^{12}$C of 3 fm and they are much less than that
of $^{184}$W of 7.4 fm.
The above means that the dropping $K_1(1270)^+$ mass scenario will lead to an enhancement (see below)
of the yield of the $K_1(1270)^+$ mesons in $K^-A$ reactions at near-threshold incident beam momenta
of interest both due to in-medium shift of the elementary production threshold to lower energy and due to
their weaker absorption in the nuclear matter.

\section*{3 Results and discussion}

\hspace{0.5cm} The model described above makes it possible to calculate
the absolute $K_1(1270)^+$ momentum differential
cross sections from the direct $K_1(1270)^+$ production process
in $K^-$$^{12}$C and $K^-$$^{184}$W collisions.
These cross sections were calculated according to Eq. (24) for three
adopted values of the $K_1(1270)^+$ in-medium mass shift $V_0$ at density $\rho_0$ at laboratory angles
of 0$^{\circ}$--45$^{\circ}$ and for initial antikaon momenta of 3.5, 2.8 and 2.5 GeV/c. They are presented,
respectively, in Figs. 3, 4 and 5. One can see from these figures
that the $K_1(1270)^+$ meson differential cross sections are notably
sensitive to its in-medium mass shift, mostly in the low-momentum region of 0.1--1.0 GeV/c, for both target nuclei
and for all antikaon momenta considered. Here, the differences between all calculations corresponding to
different options for the $K_1(1270)^+$ in-medium mass shift are well separated and experimentally
distinguishable. They are practically similar to each other for each target nucleus at initial antikaon momenta
considered.
Thus, for example, for incident $K^-$ and outgoing $K_1(1270)^+$ meson momenta of 3.5 and 0.5 GeV/c, respectively,
and in the case of $^{12}$C nucleus the $K_1(1270)^+$ yield is enhanced at mass shift $V_0=-20$ MeV by about
a factor of 2.2 as compared to that obtained for the shift $V_0=0$ MeV. When going from
$V_0=-20$ MeV to $V_0=-35$ MeV the enhancement factor is about 1.8.
In the case of $^{184}$W target nucleus these enhancement factors are about 3.3
and 2.0. At initial antikaon momentum of 2.8 GeV/c and the same outgoing kaon momentum of 0.5 GeV/c the corresponding
enhancement factors are similar and are about 1.9 and 1.7 as well as 1.7 and 1.4 in the cases
of $^{12}$C as well as $^{184}$W target nuclei, respectively. And for incident beam momentum of 2.5 GeV/c and final
kaon momentum of 0.5 GeV/c these enhancement factors are about 1.9 and 1.5 as well as 1.7 and 1.5 for
$^{12}$C as well as $^{184}$W nuclei, correspondingly. However, the $K_1(1270)^+$ low-momentum
(and high-momentum) production differential cross sections at beam momentum of 2.5 GeV/c
are less than those at antikaon momenta of 2.8 and
3.5 GeV/c by about of one to two orders of magnitude. Therefore, the $K_1(1270)^+$ meson differential
cross sections measurements in the near-threshold incident $K^-$ momentum region (at 2.8--3.5 GeV/c) will open
an opportunity to shed light on its possible mass shift in cold nuclear matter.
Such measurements could be performed in the future at the J-PARC Hadron Experimental Facility using the high-intensity
and high-momentum (up to 10 GeV/c) separated secondary $K^-$ beams in the designed now the K10 beamline [34, 44].

More detailed information about the sensitivity of the differential cross sections,
presented in Figs. 3, 4 and 5, to the $K_1(1270)^+$ mass shift is contained in Figs. 6, 7 and 8,
where the momentum dependence of the ratios
of these cross sections calculated for the $K_1(1270)^+$ mass shift $V_0$ at the central density $\rho_0$
to the analogous cross section as determined at $V_0=0$ MeV is shown on a linear scale for carbon and tungsten nuclei
at $K^-$ momenta of 3.5, 2.8 and 2.5 GeV/c, respectively
\footnote{$^)$ We recall that a comparison of a similar model cross section ratios with data
on $\eta^{\prime}$ meson photoproduction off carbon and niobium nuclei was employed in [45, 46] to extract
its in-medium mass shift at a central nuclear density.}$^)$
.
It should be pointed out that such relative observables are more favorable compared to those based on the
absolute cross sections for the aim of obtaining the information on particle in-medium modification, since
the theoretical uncertainties associated with the particle elementary production cross section essentially
cancel out in them. It can be seen from these figures that at the $K_1(1270)^+$ meson momenta less than 1.0 GeV/c
there are indeed substantial differences between the results obtained by using considered options for its
in-medium mass shift for both target nuclei and all adopted initial momenta. Moreover, Figs. 7 and 8 clearly show that
the relative $K_1(1270)^+$ yield experiences rather noticeable variation for the mass shift range of $V_0=0$ MeV
to -35 MeV also in the high-momentum region (at kaon momenta $\ge$ 1.0 GeV/c) for subthreshold $K^-$ momenta of
2.8 and 2.5 GeV/c. Thus, for incident antikaon momentum of 2.8 GeV/c and in the case of a carbon nucleus, the
distinctions between the ratios corresponding to zero mass shift and shift values of -20 and -35 MeV are of the
order of 30\% and 60\% in the $K_1(1270)^+$ momentum range $\sim$ 1.0--1.6 GeV/c, where the absolute cross sections
are the greatest. In a case of a tungsten target nucleus, these distinctions are about 40\% and 90\%, respectively.
At an initial beam momentum of 2.5 GeV/c and the same outgoing kaon momenta of 1.0--1.6 GeV/c, the corresponding
distinctions are larger: they are about 50\% and 90\% as well as 60\% and 220\% in the cases of $^{12}$C as well
as $^{184}$W nuclei, respectively. At the same time, as the antikaon beam momentum increases to 3.5 GeV/c, the
sensitivity of the high-momentum parts of the ratios (and absolute momentum distributions) considered to variations in the $K_1(1270)^+$ in-medium mass shift becomes somewhat lower. Thus, at this incident momentum and the same final kaon momenta as above, the respective distinctions, as follows from Fig. 6, are smaller (but yet are measurable) than at beam momenta of 2.8 and 2.5 GeV/c: they are about 15\% and 30\% as well as 20\% and 40\% for $^{12}$C as well
as $^{184}$W nuclei, correspondingly. This means, accounting for the above considerations, that the measurements
of the $K_1(1270)^+$ meson momentum distributions (absolute and relative) might permit to shed light also on the
momentum dependence of its in-medium mass shift (or of its effective scalar potential $V_0$) at saturation density
$\rho_0$. It should be noticed that the ratios of differential cross sections for $K_1(1270)^+$ meson production
on $^{184}$W nucleus by 3.5 and 2.8 GeV/c antikaons, shown in Figs. 6 and 7, respectively, exhibit dips at
momenta $\sim$ 0.4--0.5 GeV/c for adopted mass shift values of -20 and -35 MeV. This can be explained by the fact
that the above cross sections "are bent down" at these momenta (see Figs. 3 and 4) due to the off-shell kinematics
of the direct $K^-p$ collision and the role played by the nucleus-related effects such as the target proton
binding and Fermi motion, encoded in the nuclear spectral function $P_A({\bf p}_t,E)$. The spectral functions
for $^{12}$C and $^{184}$W, used in our calculations, are different [37--39].

The sensitivity of the $K_1(1270)^+$ meson production differential cross sections to
its in-medium mass shift $V_0$, shown in Figs. 3, 4, 5 and 6, 7, 8, can also be studied from
the measurements of the total cross sections for $K_1(1270)^+$ production in ${K^-}^{12}$C
and  ${K^-}^{184}$W collisions at the threshold incident $K^-$ momenta. Such cross sections, calculated
by integrating Eq. (24) over the $K_1(1270)^+$ momentum $p_{K_1}$
at 2.5, 2.8 and 3.5 GeV/c antikaon momenta in the low-momentum region (0.1--1.0 GeV/c)
and in the full-momentum region allowed for given beam momentum are shown in Fig. 9 as functions of the mass shift $V_0$.
It is seen from this figure that again the low-momentum region of 0.1--1.0 GeV/c shows the highest sensitivity
to this shift. Thus, for example, the ratios between the total cross sections for $K_1(1270)^+$ production by
2.5, 2.8, 3.5 GeV/c $K^-$ mesons on $^{12}$C and $^{184}$W nuclei in this momentum region, calculated with
the shift $V_0=-35$ MeV, and the same cross sections as those obtained with $V_0=0$ MeV, are
about 2.1, 1.9, 1.6 and 2.3, 2.3, 1.8, respectively.
Whereas the same ratios in the full-momentum regions are only somewhat smaller: they are about
2.0, 1.6, 1.2 for $^{12}$C and 2.2, 2.0, 1.4 for $^{184}$W, correspondingly.
The highest sensitivity of the total cross sections for $K_1(1270)^+$ production
in the low-momentum and full-momentum ranges under consideration
to the mass shift $V_0$ is observed, as one would expect, at beam momentum of 2.5 GeV/c.
However, the cross sections at this momentum are small and they are less than those at
$K^-$ momenta of 2.8 and 3.5 GeV/c by about of one and two orders of magnitude, respectively.
Since the latter ones have a measurable strength
$\sim$ 0.01--25 ${\rm \mu}$b, the total cross section measurements of $K_1(1270)^+$ meson production on
nuclei both in the low-momentum (0.1--1.0 GeV/c) and in the full momentum regions for
incident antikaon momenta not far from threshold (for momenta $\sim$ 2.8--3.5 GeV/c) with the aim of obtaining
information on its mass shift in nuclear medium look quite promising as well.

The above findings of Fig. 9 that both the low-momentum (0.1--1.0 GeV/c) and the full-momentum regions reveal a definite sensitivity to the $K_1(1270)^+$ in-medium mass shift $V_0$ at saturation density $\rho_0$
for incident $K^-$ momenta of interest are nicely supported also by the results presented in Fig. 10.
Here, the ratios of the $K_1(1270)^+$ meson production total cross sections calculated for its
mass shift $V_0$ and given in Fig. 9 to the analogous cross sections determined at $V_0=0$ MeV are shown
as functions of this mass shift.
It should be mentioned that an analysis of these ratios, as in the case of those shown in Figs. 6--8,
has the advantage that the theoretical uncertainties associated with the elementary cross section
for particle production substantially cancel out in them.
It is clearly seen from this figure that the highest sensitivity of the ratios in
both considered kinematic ranges to the mass shift $V_0$ is indeed observed at far
subthreshold antikaon momentum of 2.5 Gev/c, corresponding to the free space excess energy of -164 MeV.
For instance, at this momentum and for these
ranges the cross section ratios for $V_0=-20$ MeV are about 1.5 for both target nuclei.
At this mass shift they are similar to those at momentum of 2.8 GeV/c. The latter ones are about
1.4 and 1.5 for both considered kinematic ranges for $^{12}$C and $^{184}$W, respectively.
As the $K^-$ momentum increases to 3.5 GeV/c, the sensitivity of the total cross section ratios to
changes in the $K_1(1270)^+$ in-medium mass shift $V_0$ decreases.
Thus, in the case where $K_1(1270)^+$ mesons of
momenta of 0.1--1.0 GeV/c are produced by 3.5 GeV/c antikaons incident on $^{12}$C and $^{184}$W
targets, the ratios being considered for $V_0=-20$ MeV take smaller but yet measurable values of about 1.3
and 1.4, respectively. The analogous ratios for the production of the $K_1(1270)^+$
mesons in the full-momentum ranges by 3.5 GeV/c $K^-$ mesons in $^{12}$C and $^{184}$W nuclei are
somewhat yet smaller: they are about 1.1 and 1.2, respectively.

Therefore, one can conclude that a comparison of the low-momentum and full-momentum "integral"
results shown in Figs. 9, 10 with the respective precise experimental data, which could be also obtained
in future experiments at J-PARC [34, 44] employing near-threshold $K^-$ beams,
should allow to distinguish between possible weak ($V_0$ $\sim$ -20 MeV) and relatively weak
($V_0$ $\sim$ -35 MeV) $K_1(1270)^+$ meson mass shifts in nuclear matter, considered in the present work.

   Finally, accounting for the above considerations, we can conclude that the
$K_1(1270)^+$ differential and total cross section measurements in ${K^-}A$ reactions at initial
momenta not far from threshold (at momenta $\sim$ 2.8--3.5 GeV/c)
both in the $K_1(1270)^+$ meson low-momentum (0.1--1.0 GeV/c) and in its full-momentum ranges
will allow us to shed light on the possible $K_1(1270)^+$ in-medium mass shift.

\section*{4 Summary}

\hspace{0.5cm} Our present study was aimed at studying the possibility of extracting information about
the predicted in-medium change in the $K_1(1270)^+$ meson mass. For this,
we calculated the absolute differential and total cross
sections for the production of $K_1(1270)^+$ mesons on $^{12}$C and $^{184}$W target nuclei at
laboratory angles of 0$^{\circ}$--45$^{\circ}$ by $K^-$ mesons with momenta of 2.5, 2.8 and 3.5 GeV/c,
which are close to the threshold momentum (2.95 GeV/c) for $K_1(1270)^+$ meson production off the free
target proton at rest. These calculations have been performed within a nuclear spectral function approach,
which describes incoherent direct $K_1(1270)^+$ meson production in $K^-$ meson--proton
${K^-}p \to {K_1(1270)^+}\Xi^-$ production processes and accounts for three different options for its
in-medium mass shift (or for its effective scalar potential) at central density $\rho_0$.
We show that the differential and total (absolute and relative) $K_1(1270)^+$ antikaon-induced
production cross sections at initial momenta not far from threshold -- at momenta $\sim$ 2.8--3.5 GeV/c, at which
there is yet no strong drop in their strength, reveal
a distinct sensitivity to changes in the in-medium shift of the $K_1(1270)^+$ mass, studied in the paper,
both in the $K_1(1270)^+$ meson low-momentum (0.1--1.0 GeV/c) and in its full-momentum ranges.
This would permit evaluating this shift. Experimental data necessary for this aim can be obtained
in a dedicated experiment at the J-PARC Hadron Experimental Facility.
\\
\\

\end{document}